\documentclass[a4paper]{report}
\usepackage[utf8]{inputenc}
\usepackage[T1]{fontenc}
\usepackage{RJournal}
\usepackage{amsmath,amssymb,array}
\usepackage{booktabs}

\providecommand{\tightlist}{%
  \setlength{\itemsep}{0pt}\setlength{\parskip}{0pt}}

\newlength{\cslhangindent}
\newlength{\csllabelwidth}
\newlength{\cslentryspacingunit} 
  {}%
  {\par}
\newenvironment{CSLReferences}[2] 
 {
  \setlength{\parindent}{0pt}
  \ifodd #1
  \let\oldpar\par
  \def\par{\hangindent=\cslhangindent\oldpar}
  \fi
  \setlength{\parskip}{#2\cslentryspacingunit}
 }%
 {}
\usepackage{calc}

\begin{document}

\sectionhead{An interactive version of this article can be found at The R Journal}
\volume{15}
\volnumber{4}
\year{2023}
\month{December}

\begin{article}
\title{SUrvival Control Chart EStimation Software in R: the success Package}
\author{by Daniel Gomon, Marta Fiocco, Hein Putter, and Mirko Signorelli}

\maketitle

\abstract{%
Monitoring the quality of statistical processes has been of great importance, mostly in industrial applications. Control charts are widely used for this purpose, but often lack the ability to monitor survival outcomes. Recently, inspecting survival outcomes has become of interest, especially in medical settings where outcomes often depend on risk factors of patients. For this reason many new survival control charts have been devised and existing ones have been extended to incorporate survival outcomes. The package \texttt{success} allows users to construct risk-adjusted control charts for survival data. Functions to determine control chart parameters are included, which can be used even without expert knowledge on the subject of control charts. The package allows to create static as well as interactive charts, which are built using \texttt{ggplot2} (Wickham 2016) and \texttt{plotly} (Sievert 2020).
}

\textcolor{red}{\textbf{ORIGINAL ARTICLE}\\
Please cite this paper as:\\
D. Gomon, M. Fiocco, H. Putter, and M. Signorelli, SUrvival Control Chart EStimation Software in R: the success Package, The R Journal, vol. 15, no. 4, pp. 270–291, Jul. 2023, doi: 10.32614/rj-2023-095.\\
The original version of this manuscript is freely accessible \hyperref{https://journal.r-project.org/articles/RJ-2023-095/}{}{}{from The R Journal website}.}

\hypertarget{introduction}{%
\section{Introduction}\label{introduction}}

Inspecting the quality of a survival process is of great importance, especially in the medical field. Many of the methods currently used to inspect the quality of survival processes in a medical setting, such as funnel plots (Spiegelhalter 2005), Bernoulli Cumulative sum (CUSUM) charts (Steiner et al. 2000) and exponentially weighted moving average (EWMA) charts (Cook, Coory, and Webster 2011) work only with binary outcomes, and are thus not appropriate for survival outcomes. These charts require the continuous time outcome to be dichotomized, often leading to delays when trying to detect problems in the quality of care. To overcome this limitation, (Biswas and Kalbfleisch 2008) developed a continuous time CUSUM procedure (which we call the BK-CUSUM), that can be used to inspect survival outcomes in real time. (Gomon et al. 2022) proposed a generalization of the BK-CUSUM chart called the Continuous Time Generalized Rapid Response CUSUM (CGR-CUSUM). The CGR-CUSUM allows to estimate some of the parameters involved in the construction of the chart, overcoming the need for the user to correctly specify parameters that the BK-CUSUM procedure relies on. Recently other procedures for the continuous time inspection of survival outcomes have been developed, such as the improved Bernoulli CUSUM (Keefe et al. 2017), uEWMA chart (Steiner and Jones 2009) and STRAND chart (Grigg 2018). To the best of our knowledge there are no publicly available software implementations of these methods.

When constructing control charts in continuous time, not only the time to failure of a subject is of interest, but also the information provided by the survival up until current time is crucial. Many quality control methods cannot incorporate continuous time (survival) outcomes, requiring the continuous time outcome to be dichotomized (e.g.~30-day survival). The resulting binary data is called discrete time data. We provide an overview of some of the existing \texttt{R} packages which can be used for constructing control charts. The package \CRANpkg{qcc} (Scrucca 2004) for discrete time data contains functions to construct many types of Shewhart, binary CUSUM, EWMA and other charts. The packages \CRANpkg{qcr} (Flores 2021), \CRANpkg{qicharts} (Anhoej 2021) and \CRANpkg{ggQC} (Grey 2018) also allow for the construction of discrete time control charts, but differ in their graphical possibilities and their intended application area such as medicine, industry and economics. It is possible to assess some of the discrete time control charts by means of their zero/steady state average run length using the packages \CRANpkg{spc} (Knoth 2021) and \CRANpkg{vlad} (Wittenberg and Knoth 2020). The package \CRANpkg{funnelR} (Kumar 2018) allows for the construction of funnel plots for proportion data, a method often used in medical statistics to visualise the difference in proportion over a time frame. The package \CRANpkg{cusum} (Hubig 2019) can be used to monitor hospital performance using a Bernoulli CUSUM, also allowing users to easily determine control limits for continuously inspecting hospitals. The package however requires the input to be presented in a binary format.

Whereas many packages exist allowing for the construction of quality control charts on discrete time data, to the best of our knowledge there are currently few statistical software packages allowing for the construction of quality control charts on survival data and no R packages allowing for the continuous time inspection of survival outcomes.

The main contribution of this article is to present the R package \CRANpkg{success} (SUrvival Control Chart EStimation Software), a tool for constructing quality control charts on survival data. With this package, we aim to fill the gap in available open source software for the construction of control charts on survival data. The primary goal of \CRANpkg{success} is to allow users to easily construct the BK- and CGR-CUSUM; moreover, \CRANpkg{success} can also be used to construct the discrete time funnel plot (Spiegelhalter 2005) and the Bernoulli CUSUM chart (Steiner et al. 2000) on survival data without manually dichotomizing the outcomes. This way, users can determine the possible gain in detection speed by using continuous time quality control methods over some popular discrete time methods.

The article is structured as follows. In Section \protect\hyperlink{sec:TheoryandModels}{Theory and models} we briefly describe the theory that underlines the control charts presented in the package. Section \protect\hyperlink{sec:Rpkgsuccess}{The R package success} shows how to prepare the input data and describes the main functions and their arguments. For the reader not interested in technical details, a helper function is described in Section \protect\hyperlink{sec:parassistfct}{The parameter-assist function}. The control charts in the package are then applied to data based on a clinical trial for breast cancer in the \protect\hyperlink{sec:application}{Application} section. Finally, the article ends with a \protect\hyperlink{discussion}{Discussion} about the methods presented.

\hypertarget{sec:TheoryandModels}{%
\section{Theory and models}\label{sec:TheoryandModels}}

Throughout this article, we are interested in comparing institutional (hospital) performance for survival after a medical procedure (surgery). Even though our focus is on medical applications, the methods can be applied to any data set containing survival outcomes.

In this section we introduce the funnel plot, Bernoulli CUSUM, BK-CUSUM and CGR-CUSUM implemented in the package \CRANpkg{success}. The goal of each of the methods is to detect a deviation from a certain target performance measure and discover hospitals with increased mortality rates as quickly as possible.

\hypertarget{sec:ReadingGuide}{%
\subsubsection{Reading guide}\label{sec:ReadingGuide}}

This section consists of two parts. Section \protect\hyperlink{sec:Terminology}{Terminology} summarizes the minimal required knowledge in layman's terms. Section \protect\hyperlink{sec:MathematicalNotation}{Mathematical notation} delves into the mathematical details and assumptions. Further sections introduce the mathematical theory of the methods available in \CRANpkg{success}.

\hypertarget{sec:Terminology}{%
\subsection{Terminology}\label{sec:Terminology}}

We assume that hospitals/units have a constant and steady stream of patients/subjects coming in for a treatment of interest (e.g.~surgery). In survival quality control, we are interested in determining whether failure/death rates after treatment at a certain hospital deviate significantly from a certain target measure. A target measure defines the acceptable failure rate. This measure can be set or estimated from a large set of (historical) data. Most of the time, we are only interested in detecting an increase in failure rate as this indicates that the hospital in question is performing worse than expected, and that corrective interventions may be necessary to improve the quality of care at such a hospital.

The process is defined to be \emph{in-control} when failures occur according to the target level. It is \emph{out-of-control} if failures happen at a higher rate then expected. Hospitals may have in-control periods followed by out-of-control periods. The goal is to detect when observations at a hospital start going out-of-control as soon as possible, so action can be taken quickly.

To continuously inspect the quality of the process, we construct a control chart to monitor the process failure rate from the start of the study. Some charts only change value when an outcome is observed (discrete time), while others change value at each time point (continuous time). When the control chart exceeds a pre-defined \emph{control limit}, a signal is produced, indicating that the hospital in question is performing worse/better than the target measure. The time it takes for a control chart to exceed the control limit is called the \emph{run length} of the chart. For some control charts it is necessary to fix the expected increase in failure rate in advance. This is done by specifying a parameter \(\theta\), where \(e^\theta\) indicates the expected increase in the failure odds (discrete time) or expected increase in the failure rate (continuous time).

Not all patients have an equal probability of failure at any given point in time. For example, people who smoke may have a larger risk of failure than non-smokers. Therefore, patient's risk can be incorporated into the control charts by using a risk-adjustment model. Relevant characteristics (called covariates) are then used to determine the increase/decrease in the risk of failure for each patient.

\hypertarget{sec:MathematicalNotation}{%
\subsection{Mathematical notation}\label{sec:MathematicalNotation}}

Consider a single hospital. For each patient (\(i = 1,2,...\)) let \(A_i\) and \(X_i\) be the chronological entry time and survival/failure time from the time of entry respectively. The chronological time of failure is then given by \(T_i = A_i + X_i\). Assume that patients arrive (enter the study) at the hospital according to a Poisson process with rate \(\psi\), and that each patient \(i\) has a set of \(p\) covariates \(\mathbf{Z}_i\). Let \(h_i(x) = h_0(x) e^{\mathbf{Z}_i \pmb{\beta}}\) be the subject-specific hazard rate obtained from the Cox proportional hazards model (Cox 1972). Let \(Y_i(t) = \mathbb{1} \{ A_i \leq t \leq \min(T_i, R_i) \}\) indicate whether a patient is \emph{at risk} (of failure) at time \(t\), where \(R_i\) denotes the right censoring time of patient \(i\).

Define by \(N_i(t)\) the counting process indicating whether patient \(i\) experiences a failure at or before time \(t\). Let \(N(t) = \sum_{i \geq 1} N_i(t)\) be the total number of observed failures at the hospital at or before time \(t\). Define the cumulative intensity of patient \(i\) as \(\Lambda_i(t) = \int_0^t Y_i(u) h_i(u) du\) and \(\lambda_i(t) = Y_i(t) h_i(t)\). Note that \(\lambda_i(t)\) is equal to zero when patient \(i\) is not at risk. Similarly, let \(\Lambda(t) = \sum_{i \geq 1} \Lambda_i(t)\) be the total cumulative intensity at the hospital at time \(t\).

For some methods, we will be interested in detecting a fixed increase in the cumulative intensity at the hospital from \(\Lambda(t)\) to \(\Lambda^\theta(t) := e^\theta \Lambda(t)\). In a similar fashion, denote \(h_i^\theta := e^\theta h_i(t)\). The corresponding density and distribution functions are denoted by \(f_i^\theta\) and \(F_i^\theta\). We call \(e^\theta\) the \emph{true hazard ratio}. When \(e^\theta = 1\) (similarly \(\theta = 0\)), we say that the failure rate is \emph{in-control}. Alternatively, when \(e^\theta > 1\) we say that the failure rate is \emph{out-of-control}.

For a chart \(K(t)\) with changing values over time define the \emph{average run length} for a given control limit \(h\) as \(\mathbb{E}[\tau_h]\), where \(\tau_h = \inf\{t > 0: K(t) \geq h\}\).

\hypertarget{sec:FunnelPlot}{%
\subsection{Funnel plot}\label{sec:FunnelPlot}}

The risk-adjusted funnel plot (Spiegelhalter 2005) is a graphical method used to compare performance between hospitals over a fixed period of time.
The general structure of the data is as follows: there are k centers/hospitals (j=1\ldots k) with \(n_j\) treated patients in hospital \(j\). For each patient we observe a binary variable \(X_{i,j}\):
\begin{equation}
X_{i,j} = \begin{cases}
1, & \text{if patient } i \text{ at hospital } j \text{ had an adverse event within $C$ days, } \\
0, & \text{if patient } i \text{ at hospital } j \text{ otherwise.}
\end{cases}
\label{eq:DiscreteOutcome}
\end{equation}
We model \(X_{i,j} \sim \mathrm{Ber}(p_j)\), with \(p_j\) the probability of failure at hospital \(j\) within \(C\) days.
Consider the hypotheses:
\begin{align}
H_0: p_j = p_0 && H_1: p_j \neq p_0
\label{eq:FunnelHypotheses}
\end{align}
with \(p_0\) some baseline (in-control) failure proportion. The proportion of failures observed at hospital \(j\) is then given by \(\gamma_j = \frac{\sum_{i=1}^{n_j}X_{i,j}}{n_j }\). The asymptotic distribution of \(\gamma_j\) under \(H_0\) is \(\left. \gamma_j \right|_{H_0} \sim \mathcal{N}\left( p_0, \frac{p_0(1-p_0)}{n_j} \right)\). We can then signal an increase or decrease in the failure proportion of hospital \(j\) with confidence level \(1-2\alpha\) when
\begin{align}
\gamma_j \notin \left[  p_0 + \xi_{\alpha} \sqrt{\frac{p_0(1-p_0)}{n_j}}, p_0 - \xi_{\alpha} \sqrt{\frac{p_0(1-p_0)}{n_j}}  \right],
\label{eq:FunnelPredictionInterval}
\end{align}
with \(\xi_{\alpha}\) the \(\alpha-\)th quantile of the standard normal distribution.

It is often desirable to determine patient specific failure probabilities using some of their characteristics. A risk-adjusted funnel plot procedure can then be performed by modelling the patient specific failure probability using a logistic regression model: \(p_i = \frac{1}{1 + e^{-\beta_0 + \pmb{Z}_i \pmb{\beta}}}\), where \(\pmb{Z}_i\) is the vector of \(p\) covariates for patient \(i\). The expected number of failures at hospital \(j\) is then given by
\begin{align}
E_j &= \mathbb{E} \left[ \sum_{i=1}^{n_j} X_{j,i} \right] =  \sum_{i=1}^{n_j} p_i =  \sum_{i=1}^{n_j} \frac{1}{1 + e^{-\beta_0 + \pmb{Z}_i \beta}}.
\label{eq:RAExpectedDeaths}
\end{align}
Let \(O_j\) be the observed number of failures at hospital \(j\), the risk-adjusted proportion of failures at hospital \(j\) is given by \(\gamma_{j}^{\mathrm{RA}} = \frac{O_j}{E_j} \cdot p_0\). The quantity \(\gamma_{j}^{\mathrm{RA}}\) is then used in Equation \eqref{eq:FunnelPredictionInterval} instead of \(\gamma_j\).

The funnel plot can be used to compare hospital performance over a fixed time period. The funnel plot is often used for monitoring the quality of a process by repeatedly constructing funnel plots over different time intervals. We advocate against such an inspection scheme, as it introduces an increased probability of a type I error due to multiple testing. We recommend to only use the funnel plot as a graphical tool to visually inspect the proportion of failures at all hospitals over a time frame. The funnel plot is a discrete time method and can therefore only be used to compare overall performance over a time span. To determine whether a hospital was performing poorly during the time of interest, one of the following CUSUM charts should be used.

\hypertarget{BernoulliCUSUM}{%
\subsection{Bernoulli cumulative sum (CUSUM) chart}\label{BernoulliCUSUM}}

The Bernoulli CUSUM chart (Steiner et al. 2000) can be used to sequentially test whether the failure rate of patients at a single hospital has changed starting from some patient \(\nu \geq 1\). Consider a hospital with patients \(i = 1, ..., \nu, ...\) and a binary outcome:
\begin{align}
X_{i} = \begin{cases}
1, & \text{if patient } i \text{ had an undesirable outcome within $C$ days,} \\
0, & \text{if patient } i  \text{ had a desirable outcome within $C$ days.}
\end{cases}
\label{eq:BernoulliOutcome}
\end{align}
We model \(X_i \sim \mathrm{Ber}(p_i)\) with \(p_i\) the failure probability within \(C\) days for patient \(i\). The Bernoulli CUSUM can be used to test the hypotheses of an increased failure rate starting from some patient \(\nu\):
\begin{equation}
\begin{aligned}
H_0: &X_1, X_2, ... \sim \text{Ber}(p_0)  & &H_1: 
\begin{array}{l}
X_1, ..., X_{\nu-1} \sim \text{Ber}(p_0)\\
X_{\nu},X_{\nu+1}, .... \sim \text{Ber}(p_1)
\end{array},
\end{aligned}
\label{eq:BernoulliHypotheses}
\end{equation}
where \(\nu \geq 1\) is not known in advance, \(p_0 < p_1\), and patient outcomes are ordered according to the time of entry into the study \(A_i\).

The Bernoulli CUSUM statistic is given by:
\begin{align}
S_n = \max \left( 0, S_{n-1} + W_n   \right),
\label{eq:BernoulliStatistic}
\end{align}
with \(W_n = X_n \ln \left( \frac{p_1(1-p_0)}{p_0(1-p_1)} \right) + \ln \left( \frac{1-p_1}{1-p_0} \right)\). Alternatively, it is possible to reformulate the chart in terms of the Odds Ratio \(OR = \frac{p_1(1-p_0)}{p_0(1-p_1)} =: e^\theta\). In that case, \(W_n = X_n \ln \left( e^\theta \right) + \ln \left( \frac{1}{1-p_0 + e^\theta p_0} \right)\). The null hypothesis is rejected when the value of the chart exceeds a control limit \(h\).

A risk-adjusted procedure may be performed by modelling patient-specific failure probability (\(p_{0,i}\)) using a logistic regression model. The risk-adjusted Bernoulli CUSUM can be used as a sequential quality control method for binary outcomes. Dichotomizing the outcome (survival time) can lead to delays in detection. When survival outcomes are available, it can therefore be beneficial to construct one of the CUSUM charts described in the following sections.

\hypertarget{sec:BKKCUSUM}{%
\subsection{Biswas and Kalbfleisch CUSUM (BK-CUSUM)}\label{sec:BKKCUSUM}}

The BK-CUSUM chart can be used to continuously test whether the failure rate of patients at the hospital has changed at some point in time. Consider a hospital with patients \(i = 1,2, ...\) and assume that the patient specific hazard rate is given by \(h_i(x) = h_0(x) e^{\mathbf{Z}_i \beta}\). The BK-CUSUM chart can be used to test the hypotheses that the baseline hazard rate of all active patients has increased from \(h_0(x)\) to \(h_0(x) e^{\theta_1}\) at some point in time \(s > 0\) after the start of the study:
\begin{equation}
\begin{aligned}
H_0: X_i \sim \Lambda_i(t), i = 1,2,...  & &H_1: 
\begin{array}{l}
X_i \sim \left. \Lambda_i(t) \right| t < s , i = 1, 2, ... \\
X_i \sim \left. \Lambda_i^{\theta_1}(t) \right| t \geq s, i = 1, 2, ... 
\end{array},
\end{aligned}
\label{eq:BKhypotheses}
\end{equation}
where \(\theta_1 > 0\) is the user's estimate of the true hazard ratio \(\theta\) and \(s > 0\) is the unknown time of change in hazard rate.

The likelihood ratio chart associated with the hypotheses in \eqref{eq:BKhypotheses} is given by:
\begin{align}
BK(t) &= \max_{0 \leq s \leq t} \left\lbrace \theta_1 N(s,t) - \left(  e^{\theta_1} - 1 \right)  \Lambda(s,t) \right\rbrace,
\label{eq:BKstatistic}
\end{align}
where the estimated hazard ratio \(e^{\theta_1} > 1\) has to be prespecified, \(N(s,t) = N(t) - N(s)\) and \(\Lambda(s,t) = \Lambda(t) - \Lambda(s)\). The null hypothesis is rejected when the value of the chart exceeds the control limit.

The BK-CUSUM chart can lead to faster detection speeds than the Bernoulli CUSUM chart as the hypotheses can be tested at any point in time, rather than just at the (dichotomized) times of outcome. Unfortunately the chart requires users to specify \(\theta_1\) to estimate \(\theta\), which is not known a priori in most practical applications. Misspecifying this parameter can lead to large delays in detection (Gomon et al. 2022).

\hypertarget{sec:CGRCUSUM}{%
\subsection{Continuous time Generalized Rapid response CUSUM (CGR-CUSUM)}\label{sec:CGRCUSUM}}

The CGR-CUSUM chart can be used to test the following hypotheses:
\begin{equation}
\begin{aligned}
H_0: X_i \sim \Lambda_i, i = 1,2,...  & &H_1: 
\begin{array}{l}
X_i \sim \Lambda_i, i = 1, 2, ..., \nu -1 \\
X_i \sim \Lambda_i^\theta, i = \nu, \nu +1, ... .
\end{array},
\end{aligned}
\label{eq:CGRhypotheses}
\end{equation}
where \(e^\theta\) and \(\nu\) do not need to be prespecified. The CGR-CUSUM chart is then given by
\begin{align}
        CGR(t) &= \max_{1 \leq \nu \leq n} \left\lbrace \hat{\theta}_{\geq \nu}(t) N_{\geq \nu}(t) - \left( \exp\left(\hat{\theta}_{\geq \nu}(t)\right) - 1 \right)  \Lambda_{\geq \nu}(t) \right\rbrace,
        \label{eq:CGRStatistic}
\end{align} where the subscript ``\(\geq \nu\)'' stands for all subjects after the potential change point \(\nu\): \(N_{\geq \nu}(t) = \sum_{i \geq \nu} N_i(t), \Lambda_{\geq \nu}(t) = \sum_{i \geq \nu} \Lambda_i(t)\)
and \(\hat{\theta}_{\geq \nu}(t) = \max \left(0, \log \left( \frac{N_{\geq \nu}(t)}{ \Lambda_{\geq \nu}(t) } \right) \right)\). The null hypothesis is rejected when the value of the chart exceeds the control limit.

In contrast to the BK-CUSUM where an estimate of \(e^\theta\) had to be specified in advance, the CGR-CUSUM uses the maximum likelihood estimate \(e^{\hat{\theta}}\) to estimate the true hazard ratio \(e^\theta\) from the data. This means that when \(e^{\theta_1}\) is misspecified in the BK-CUSUM, the CGR-CUSUM can lead to quicker detections. In practice the real hazard ratio \(e^\theta\) is never known in advance and may vary over time. The maximum likelihood estimator can alleviate this problem, therefore the CGR-CUSUM is generally the preferred chart. The major difference between the two charts is that the BK-CUSUM tests for a sudden change in the failure rate of all patients currently at risk of failure, while the CGR-CUSUM tests for a sudden change in the failure rate of all patients currently at risk of failure who have entered the hospital after a certain time. A drawback of the CGR-CUSUM is that the computation of the MLE \(\hat{\theta}\) requires sufficient information (in the form of survival times/failures) to converge to the true value. This means that the chart can be unstable at the beginning of the study and might not provide reliable values for hospitals with low volumes of patients.

\hypertarget{sec:ChoosingControlLimits}{%
\subsection{Choosing control limits}\label{sec:ChoosingControlLimits}}

For the funnel plot in Section \protect\hyperlink{sec:FunnelPlot}{Funnel plot}, it is sufficient to choose a confidence level to determine which hospitals are performing worse/better than the baseline. For the CUSUM charts, instead, it is necessary to choose a control limit \(h\), so that a signal is produced when the value of the chart exceeds \(h\). The most common ways to choose this control limit are to either restrict the in-control average run length (ARL) of the chart, or to restrict the type I error over a certain time period. With the first method, one could choose to restrict the in-control ARL to approximately \(5\) years, so that on average we would expect a hospital which performs according to the target to produce a false signal (detection) once every \(5\) years. Using the second method, one could choose the control limit such that at most a proportion \(\alpha\) of the in-control hospitals yields a signal (false detection) within a period of \(5\) years. For the (risk-adjusted) Bernoulli CUSUM plenty of results exist allowing for the numerical estimation of the ARL, most of which are implemented in the packages \CRANpkg{spc} and \CRANpkg{vlad}. For continuous time control charts such results are lacking, therefore in the \CRANpkg{success} package we choose to determine control limits by restricting the type I error probability \(\alpha\) over a chosen time frame. We estimate these control limits by means of a simulation procedure.

\hypertarget{sec:Rpkgsuccess}{%
\section{The R package success}\label{sec:Rpkgsuccess}}

The package \CRANpkg{success} can be used by both laymen and experts in the field of quality control charts. In Section \protect\hyperlink{input-data}{Input data} we describe the general data structure to be used for constructing control charts by means of an example data set. In Section \protect\hyperlink{sec:parassistfct}{The parameter-assist function}, a function is described that can be used to determine all necessary parameters for the construction of control charts for the user not interested in technical details. In Sections \protect\hyperlink{sec:ManualRiskAdjustment}{Manual risk-adjustment} - \protect\hyperlink{sec:CGRCUSUMfunction}{The CGR-CUSUM function} we present the functions that can be used to compute the control charts described in Section \protect\hyperlink{sec:TheoryandModels}{Theory and models}.

\hypertarget{input-data}{%
\subsection{Input data}\label{input-data}}

All methods in this package require the user to supply a \texttt{data.frame} for the construction of control charts and the estimation of a benchmark failure rate. We show how to use the \CRANpkg{success} package by means of an enclosed data set.

\hypertarget{example-data-set}{%
\subsubsection{Example data set}\label{example-data-set}}

The data frame \texttt{surgerydat} contains \(32529\) survival times, censoring times and covariates (age, BMI and sex) of patients from \(45\) hospitals with \(15\) small, medium and large hospitals (\(0.5\), \(1\) and \(1.5\) patients per day). Patients enter the hospitals for \(2\) years (\(730\) days) after the start of the study. Survival times were generated using a risk-adjusted Cox proportional hazards model using inverse transform sampling (Austin 2012), with coefficient vector \(\beta =\) \texttt{c(\ age\ =\ 0.003,\ BMI\ =\ 0.02,\ sexmale\ =\ 0.2)} and exponential baseline hazard rate \(h_0(t, \lambda = 0.01) e^\theta\). Hospitals are numbered from \(1\) to \(45\) with hazard ratio \(\theta\) sampled from a normal distribution with mean \(0\) and standard deviation \(0.4\). This means that some hospitals are performing better or equal to baseline (\(\theta \leq 0\)) and some are performing worse (\(\theta > 0\)).

An overview of the data set can be found in Table \ref{tab:surgerydat-static}.

\begin{verbatim}
library(success)
data("surgerydat", package = "success")
head(surgerydat, 5)
\end{verbatim}

\begin{table}

\caption{\label{tab:surgerydat-static}Overview of the enclosed `surgerydat` data set.}
\centering
\begin{tabular}[t]{r|r|r|r|r|r|r|l|r}
\hline
entrytime & survtime & censorid & unit & exptheta & psival & age & sex & BMI\\
\hline
5 & 21 & 0 & 1 & 1.887204 & 0.5 & 70 & male & 29.52\\
\hline
9 & 19 & 1 & 1 & 1.887204 & 0.5 & 68 & male & 24.06\\
\hline
10 & 64 & 1 & 1 & 1.887204 & 0.5 & 101 & female & 20.72\\
\hline
20 & 64 & 1 & 1 & 1.887204 & 0.5 & 67 & female & 24.72\\
\hline
21 & 0 & 1 & 1 & 1.887204 & 0.5 & 44 & male & 27.15\\
\hline
\end{tabular}
\end{table}

Each row represents a patient. The first \(2\) columns (\texttt{entrytime} and \texttt{survtime}) are crucial for the construction of control charts. These columns have to be present in the data. The time scale of \texttt{entrytime} and \texttt{survtime} must be the same. They can both only be supplied in a \texttt{"numeric"} format (dates are not allowed). In the example data, the time scale is in days; \texttt{entrytime} is the number of days since the start of the study and \texttt{survtime} is the survival time since the time of entry. The column \texttt{censorid} is a censoring indicator, with \(0\) indicating right-censoring and \(1\) that the event has occurred. If \texttt{censorid} is missing, a column of 1's will automatically be created, assuming that no observations were right-censored. All relevant functions in the package will print a warning when this happens. The column \texttt{unit} (indicating the hospital number) is required for the construction of a funnel plot, but not for the CUSUM charts. This is because CUSUM charts are constructed separately for every unit, requiring the user to manually subset the data for each unit. The columns \texttt{exptheta} and \texttt{psival} indicate the parameters \(e^{\theta}\) and \(\psi\) used to create the simulated data set. The last three columns \texttt{age}, \texttt{sex} and \texttt{BMI} are the covariates of each individual. In a user supplied \texttt{data.frame} these can of course take any desired name.

\hypertarget{sec:parassistfct}{%
\subsection{The parameter-assist function}\label{sec:parassistfct}}

Readers not interested in technical details can use the \texttt{parameter\_assist} function to determine most of the parameters required for constructing the control charts in this package.

The \texttt{parameter\_assist} function can be used to determine control limits for all control charts described in Section \protect\hyperlink{sec:TheoryandModels}{Theory and models}. The function guides users through the following \(3\) steps:

\begin{itemize}
\tightlist
\item
  Step 1: Specify arguments to \texttt{parameter\_assist}.
\item
  Step 2: Determine control limit(s) for the required control chart(s) by feeding the output of Step 1 to one of the \texttt{*\_control\_limit} functions.
\item
  Step 3: Construct the desired control chart by feeding the output of Step 1 to the function, as well as the control limit from Step 2.
\end{itemize}

Step 2 can be skipped for the \texttt{funnel\_plot} function, as funnel plots do not require control limits.

The \texttt{parameter\_assist} function has the following arguments:

\begin{itemize}
\tightlist
\item
  \texttt{baseline\_data} \textbf{(required)}: a \texttt{data.frame} in the format described in Section \protect\hyperlink{input-data}{Input data}. It should contain the data to be used to determine the target performance metric (both for discrete and continuous time charts). This data is used to determine what the ``acceptable'' failure rate is, as well as how much risk-adjustment variables influence the probability of observing an event. Preferably this should be data of subjects that were known to fail at acceptable rates, or historical data which we want to compare with. Usually all available data is used instead, resulting in the average performance being selected as the target;
\item
  \texttt{data} \textbf{(required)}: a \texttt{data.frame} in the format described in Section \protect\hyperlink{input-data}{Input data}. It should contain the data used to construct a control chart. For this data we want to know whether it adheres to the target determined in \texttt{baseline\_data} or not. For example: the data from one hospital;
\item
  \texttt{formula} (optional): a formula indicating in which way the linear predictor of the risk-adjustment models should be constructed in the underlying generalized linear model or Cox proportional hazards model. Only the right-hand side of the formula will be used. If left empty, no risk-adjustment procedure will be performed;
\item
  \texttt{followup} \textbf{(required for discrete time charts)}: a numeric value which has to be in the same unit as \texttt{entrytime} and \texttt{survtime} and should specify how long after \texttt{entrytime} we consider the binary outcome (failure/non-failure) of the patient. This argument can be left empty when the user does not want to construct discrete time charts;
\item
  \texttt{theta} (recommended for Bernoulli and BK-CUSUM): the expected increase in the log-odds of failure/hazard rate. Default is \(\log(2)\), meaning the goal will be a detection of a doubling in the failure rate;
\item
  \texttt{time} (recommended): time interval for type I error to be determined. Default is the largest \texttt{entrytime} of subjects in \texttt{baseline\_data};
\item
  \texttt{alpha} (recommended): required type I error to control over the time frame specified in \texttt{time}. Default is \(0.05\).
\end{itemize}

The first two arguments should always be specified. Depending on the number of additional arguments specified, different functions in the package can be used. Most arguments have default values, but these may not always be suitable for the desired inspection scheme. Risk-adjusted procedures can only be constructed if at least \texttt{formula} is specified.

An example where all arguments are specified is provided below, but specifying only \texttt{baseline\_data} and \texttt{data} is sufficient to construct a CGR-CUSUM without risk-adjustment.

\begin{verbatim}
assisted_parameters <- parameter_assist(
  baseline_data = subset(surgerydat, entrytime < 365), 
  data = subset(surgerydat, entrytime >= 365 & unit == 1), 
  formula = ~age + sex + BMI, 
  followup = 30,
  theta = log(2),
  time = 365,
  alpha = 0.05)
\end{verbatim}

We use data on patients arriving in the first year (\texttt{entrytime\ \textless{}\ 365}) to determine the target performance measure. We then construct control charts on the first hospital (\texttt{unit\ ==\ 1}) using information on all patients arriving after the first year. Risk-adjustment is performed using the \(3\) available covariates. We choose to consider patient followup \(30\) days after surgery and want to detect a doubling of failure rate. The last two arguments were chosen such that the type I error of the procedure is restricted to \(0.05\) within \(1\) year (on average \(1\) in \(20\) hospitals performing according to baseline will be detected within \(1\) year).

The \texttt{parameter\_assist} function then returns a list of arguments to supply to other functions in this package:

\begin{verbatim}
names(assisted_parameters)
\end{verbatim}

\begin{verbatim}
#>  [1] "call"          "data"          "baseline_data" "glmmod"       
#>  [5] "coxphmod"      "theta"         "psi"           "time"         
#>  [9] "alpha"         "maxtheta"      "followup"      "p0"
\end{verbatim}

The user can manually feed the determined parameters to other functions in this package. Conversely, it is possible to feed the output of the \texttt{parameter\_assist} function to the following functions directly:

\begin{itemize}
\tightlist
\item
  \texttt{funnel\_plot}
\item
  \texttt{bernoulli\_control\_limit} and \texttt{bernoulli\_cusum}
\item
  \texttt{bk\_control\_limit} and \texttt{bk\_cusum}
\item
  \texttt{cgr\_control\_limit} and \texttt{cgr\_cusum}
\end{itemize}

Step 1 was performed in the code above. For Step 2 we feed the output of \texttt{parameter\_assist} to determine control limits for the Bernoulli, BK- and CGR-CUSUM charts.

\begin{verbatim}
bernoulli_control <- bernoulli_control_limit(assist = assisted_parameters)
bk_control <- bk_control_limit(assist = assisted_parameters)
cgr_control <- cgr_control_limit(assist = assisted_parameters)
\end{verbatim}

The determined control limits can then be fed to the control chart functions to finish Step 3.

\begin{verbatim}
bernoulli_assist <- bernoulli_cusum(assist = assisted_parameters, 
                                    h = bernoulli_control$h)
bk_assist <- bk_cusum(assist = assisted_parameters, h = bk_control$h)
cgr_assist <- cgr_cusum(assist = assisted_parameters, h = cgr_control$h)
\end{verbatim}

\begin{figure}
\centering
\includegraphics{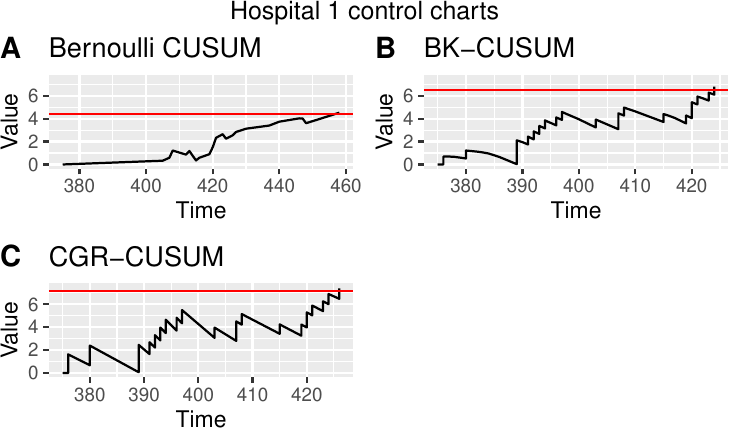}
\caption{\label{fig:assisted-plots-pdf}Bernoulli, BK- and CGR-CUSUM charts for hospital 1 in the surgery data set starting from 1 year after the start of the study.}
\end{figure}

We plot the control charts using the \texttt{plot} function (see Figure \ref{fig:assisted-plots-pdf}). The Bernoulli CUSUM jumps upward every time a failure is observed \(30\) days after patient entry and downward every time no failure is observed at that point. The BK- and CGR-CUSUM make an upward jump directly when a failure has been observed, and slope downward as long as no failures are happening. The BK- and CGR-CUSUM cross their respective control limits (the red lines) at approximately the same time after the start of the study, producing a signal. The Bernoulli CUSUM does so a little while later, producing a delayed signal.

The run length of the charts (time until control limit is reached) can then be determined using the \texttt{runlength} function.

\begin{verbatim}
runlength(bernoulli_assist, h = bernoulli_control$h)
\end{verbatim}

\begin{verbatim}
#> [1] 83
\end{verbatim}

Note that the run length of the charts are determined from the smallest time of entry of subjects into specified \texttt{data}. The study therefore starts at the moment the first subject has a surgery (in this case, at day \(375\)).

The funnel plot does not require control limits, therefore steps \(2\) and \(3\) can be skipped. We use the \texttt{plot} function to display the funnel plot.

\begin{verbatim}
funnel_assist <- funnel_plot(assist = assisted_parameters)
plot(funnel_assist, label_size = 2) + ggtitle("Funnel plot of surgerydat")
\end{verbatim}

\begin{figure}
\centering
\includegraphics{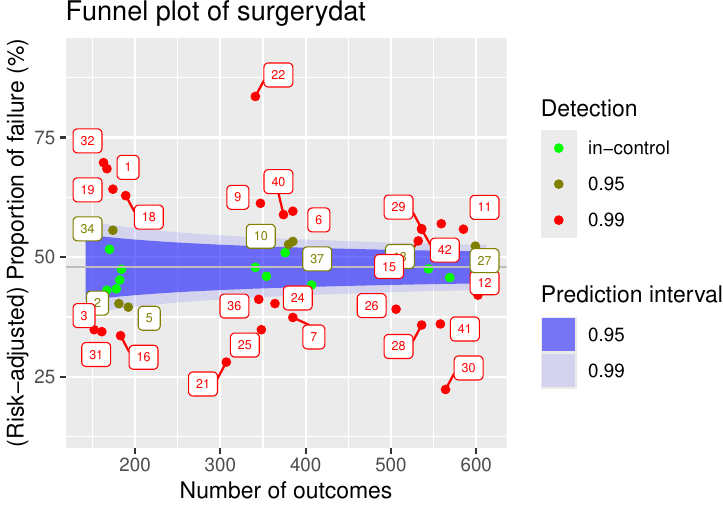}
\caption{\label{fig:assisted-funnel}Funnel plot for hospitals over the first year of the surgery data set. Target performance determined as average over the considered data.}
\end{figure}

The resulting plot can be seen in Figure \ref{fig:assisted-funnel}. The blue shaded regions indicate the \(95\) and \(99\) percent prediction intervals. Each dot represents a hospital, with the colour representing the prediction limits at which this hospital would be signaled. When a hospital falls outside of a prediction interval, it will be signaled at that level.

\hypertarget{sec:ManualRiskAdjustment}{%
\subsection{Manual risk-adjustment}\label{sec:ManualRiskAdjustment}}

Risk-adjustment models should be estimated on a data set known to have in-control failures, as this allows the coefficients to be determined as precisely as possible. In real life applications it is not known in advance which hospitals have had in-control failures. It is therefore common practice to use all available data to determine risk-adjustment models.

We consider a logistic model to use for risk-adjustment in the discrete time methods (funnel plot and Bernoulli CUSUM), using all available data of patients with surgeries in the first year of the study. We use 30 days mortality as outcome for these charts.

\begin{verbatim}
baseline_data <- subset(surgerydat, entrytime <= 365)
followup <- 30
glm_risk_model <- glm((survtime <= followup) & (censorid == 1) ~ age + sex + BMI,
                      data = baseline_data, family = binomial)
\end{verbatim}

Then we estimate a Cox proportional hazards model to use for risk-adjustment in the continuous time BK- and CGR-CUSUM, using the same baseline data. For this we use the functions \texttt{Surv} and \texttt{coxph} from the package \CRANpkg{survival} (Terry M. Therneau and Patricia M. Grambsch 2000).

\begin{verbatim}
require(survival)
coxph_risk_model <- coxph(Surv(survtime, censorid) ~ age + sex + BMI,
                          data = baseline_data)
\end{verbatim}

Conversely, we can manually specify a risk-adjustment model:

\begin{verbatim}
RA_manual <- list(formula = ~ age + sex + BMI, 
                  coefficients = c(age = 0.003, BMI = 0.02,  
                                         sexmale = 0.2)) 
\end{verbatim}

This is useful for users who do not want to use the package \CRANpkg{survival} for the estimation of the models.

\hypertarget{sec:funnelplotfunction}{%
\subsection{The funnel plot function}\label{sec:funnelplotfunction}}

The \texttt{funnel\_plot} function can be used to construct the funnel plot described in Section \protect\hyperlink{sec:FunnelPlot}{Funnel plot}. The code below constructs a funnel plot over the first \(1\) year (\texttt{ctime\ =\ 365}) on the simulated data set. By not specifying \texttt{ctime} the funnel plot is constructed over all data (\(2\) years). By leaving the parameter \texttt{p0} empty, the average failure proportion within \(30\) days is used as baseline failure probability.

\begin{verbatim}
funnel <- funnel_plot(data = surgerydat, ctime = 365, 
                      glmmod = glm_risk_model, followup = 30)
\end{verbatim}

The function creates an object of class \texttt{\textquotesingle{}funnelplot\textquotesingle{}}. The \texttt{plot} function can be used on classes in the \CRANpkg{success} package. This creates a \texttt{\textquotesingle{}gg\textquotesingle{}} object, whose graphical parameters can further be edited by the user using the package \CRANpkg{ggplot2} (Wickham 2016).

\begin{figure}
\centering
\includegraphics{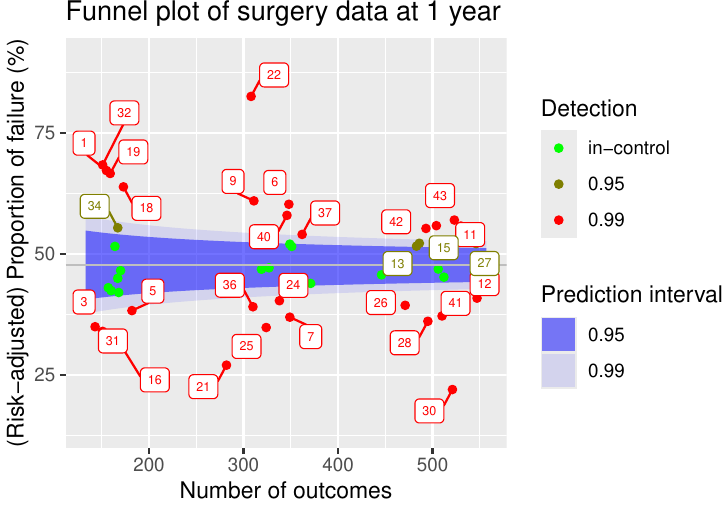}
\caption{\label{fig:funnel-p}Funnel plot of hospitals in the surgery data, using all information known after 1 year of the study.}
\end{figure}

The resulting plot can be seen in Figure \ref{fig:funnel-p}. By default, the \texttt{funnel\_plot} function will display \(95 \%\) and \(99 \%\) prediction intervals using blue shaded regions. The intervals and fill colour can be changed using the \texttt{predlim} and \texttt{col\_fill} arguments respectively. Additionally, the unit label for detected units will be displayed in the plot. In the \texttt{plot} function, the size of the labels can be adjusted using the \texttt{label\_size} argument or they can be disabled by setting the \texttt{unit\_label} argument to \texttt{FALSE}.

A summary of the funnel plot can be obtained by using the \texttt{summary} function.

\begin{verbatim}
head(summary(funnel), 5)
\end{verbatim}

\begin{table}

\caption{\label{tab:summaryfunnel-static}Summary statistics for the funnel plot.}
\centering
\begin{tabular}[t]{l|r|r|r|r|l|l}
\hline
unit & observed & expected & numtotal & p & 0.95 & 0.99\\
\hline
1 & 105 & 74.49454 & 155 & 0.6724872 & worse & worse\\
\hline
2 & 71 & 80.60038 & 168 & 0.4202816 & in-control & in-control\\
\hline
3 & 50 & 68.25903 & 143 & 0.3494854 & better & better\\
\hline
4 & 76 & 80.64515 & 167 & 0.4496292 & in-control & in-control\\
\hline
5 & 70 & 87.22662 & 182 & 0.3828848 & better & better\\
\hline
\end{tabular}
\end{table}

The resulting statistics can be found in Table \ref{tab:summaryfunnel-static}.

\hypertarget{sec:controllimitscusum}{%
\subsection{Control limits for CUSUM functions}\label{sec:controllimitscusum}}

All CUSUM charts considered in this article require control limits to signal changes in the failure rate. When the value of a chart exceeds this control limit, a signal is produced. The \CRANpkg{success} package can be used to determine control limits such that the type I error of the CUSUM procedure is restricted over some desired time frame. This is achieved by using the \texttt{*\_control\_limit} functions. We briefly discuss the underlying three steps of the simulation procedure, meanwhile explaining some key parameters that are shared across the functions.

\begin{itemize}
\tightlist
\item
  Step 1: Generate \texttt{n\_sim} units (hospitals), with subjects at each unit arriving according to a Poisson process with rate \texttt{psi} over a certain time frame specified by the parameter \texttt{time}. We therefore expect each unit to represent approximately \texttt{psi} \(\times\) \texttt{time} subjects. If specified, sample covariate values for subjects from \texttt{baseline\_data}. Generate binary outcomes/survival times for each patient according to the specified baseline (risk-adjustment) model (for details see \protect\hyperlink{sec:ManualRiskAdjustment}{Manual risk-adjustment}). Note that the simulated units represent the in-control situation.
\item
  Step 2: For each simulated unit, determine the CUSUM chart over the considered \texttt{time}.
\item
  Step 3: Determine the largest possible control limit such that at most a proportion \texttt{alpha} of the \texttt{n\_sim} in-control units would be detected using this control limit. This value then represents a simulation estimate of the control limit for a procedure with a type I error of \texttt{alpha} over \texttt{time}.
\end{itemize}

Increasing \texttt{n\_sim} will increase the accuracy of the determined control limit, at the cost of increased computation time. Similarly, increasing \texttt{time} and \texttt{psi} will also increase the computation time. These parameters however are usually determined by the underlying inspection problem. Changing \texttt{alpha} does not influence computation time.

All \texttt{*\_control\_limit} functions have a \texttt{seed} argument that can be used to obtain reproducible simulation results by setting an initial state for pseudorandom number generation. This makes sure that the procedure in Step 1 is reproducible, as the final two steps do not involve randomness. The boolean parameter \texttt{pb} can be used to display a progress bar for Step 2, which can be useful if the control limit has to be determined at a high accuracy. The \texttt{h\_precision} argument can be used to specify the required number of significant digits in determining the control limit. Choosing a high value for \texttt{h\_precision} is only useful when the constructed CUSUM charts show only very minor fluctuations or when \texttt{n\_sim} is very high, warranting a very accurate determination of the control limit. The default of two significant digits will suffice in most situations.

\hypertarget{sec:bernoullicusumfunction}{%
\subsection{The Bernoulli CUSUM function}\label{sec:bernoullicusumfunction}}

The \texttt{bernoulli\_cusum} function can be used to construct the Bernoulli CUSUM detailed in Section \protect\hyperlink{BernoulliCUSUM}{Bernoulli cumulative sum (CUSUM) chart}. The Bernoulli CUSUM uses the same dichotomized outcome as the funnel plot. For this reason, the syntax of \texttt{bernoulli\_cusum} is quite similar to that of \texttt{funnel\_plot}. In this section we will construct a Bernoulli CUSUM for the ninth hospital in the simulated data set, again using \(30\) day post operative survival as outcome and aiming to detect an increase of the odds ratio to \(2\).

\hypertarget{determining-control-limits}{%
\subsubsection{Determining control limits}\label{determining-control-limits}}

The Bernoulli CUSUM produces a signal when the value of the chart exceeds a value \(h\) called the control limit. The \texttt{bernoulli\_control\_limit} function can be used to determine a control limit such that the type I error of the Bernoulli CUSUM procedure is restricted over some desired time frame. Suppose we want to restrict the type I error of the procedure to \(0.05\) over the time frame of \(1\) year at a hospital with an average of \(1\) patient per day undergoing surgery. We determine the control limit as follows:

\begin{verbatim}
bern_control <- bernoulli_control_limit(
  time = 365, alpha = 0.05, followup = 30, psi = 1, 
  glmmod = glm_risk_model, baseline_data = surgerydat, theta = log(2))
\end{verbatim}

The determined control limit \(h\) can then be retrieved by:

\begin{verbatim}
bern_control$h
\end{verbatim}

\begin{verbatim}
#> [1] 5.56
\end{verbatim}

By default, the control limit is determined using \(200\) simulated units (hospitals). As the Bernoulli CUSUM is not very computationally intensive, it is usually possible to determine the control limit with higher precision by increasing the \texttt{n\_sim} argument.

\hypertarget{sec:constructingthechart}{%
\subsubsection{Constructing the chart}\label{sec:constructingthechart}}

After determining the control limit, we can construct the control chart. The \texttt{bernoulli\_cusum} function requires the user to specify one of the following combinations of parameters:

\begin{itemize}
\tightlist
\item
  \texttt{glmmod} \& \texttt{theta}
\item
  \texttt{p0} \& \texttt{theta}
\item
  \texttt{p0} \& \texttt{p1}
\end{itemize}

\begin{figure}
\centering
\includegraphics{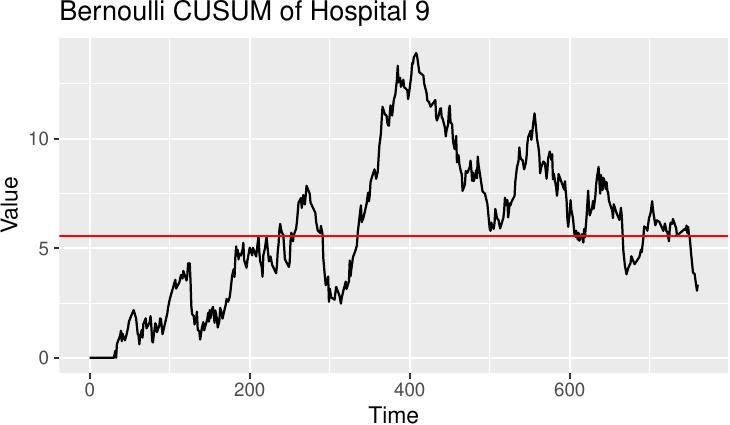}
\caption{\label{fig:Bernoulli2}Bernoulli CUSUM for hospital 9 over the study duration. Target performance estimated from the failure rate of patients treated in the first year of the study.}
\end{figure}

Only the first option allows for risk-adjustment. The difference between these parametrizations is described in Section \protect\hyperlink{BernoulliCUSUM}{Bernoulli cumulative sum (CUSUM) chart}. We construct the Bernoulli CUSUM on the data of the ninth simulated hospital, aiming to detect whether the odds ratio of failure for patients is \(2\) (\texttt{theta\ =\ log(2)}). Using the \texttt{plot} function we obtain a \texttt{\textquotesingle{}gg\textquotesingle{}} object (see Figure \ref{fig:Bernoulli2}).

\begin{verbatim}
Bernoulli <- bernoulli_cusum(
  data = subset(surgerydat, unit == 9),
  glmmod = glm_risk_model, followup = 30, theta = log(2))
plot(Bernoulli, h = bern_control$h) + 
  ggtitle("Bernoulli CUSUM of Hospital 9")
\end{verbatim}

The run length of the chart, in Figure \ref{fig:Bernoulli2} visible as the time at which the chart first crosses the red line, can be found using the \texttt{runlength} function.

\hypertarget{sec:BKCUSUMfunction}{%
\subsection{The BK-CUSUM function}\label{sec:BKCUSUMfunction}}

The \texttt{bk\_cusum} function can be used to construct the BK-CUSUM chart presented in Section \protect\hyperlink{sec:BKKCUSUM}{Biswas and Kalbfleisch CUSUM (BK-CUSUM)}. The chart is no longer constructed using dichotomized outcomes, therefore leading to faster detections on survival data than discrete time methods.

\hypertarget{determining-control-limits-1}{%
\subsubsection{Determining control limits}\label{determining-control-limits-1}}

The BK-CUSUM produces a signal when the value of the chart exceeds a value \(h\) called the control limit. The \texttt{bk\_control\_limit} function can be used to determine a control limit such that the type I error of the BK-CUSUM procedure is restricted over some desired time frame. Suppose we want to restrict the type I error of the procedure to \(0.05\) over the time frame of \(1\) year for a hospital with an average of \(1\) patient per day undergoing surgery. The control limit is determined as follows:

\begin{verbatim}
BK_control <- bk_control_limit(
  time = 365, alpha = 0.05, psi = 1, coxphmod = coxph_risk_model,
  baseline_data = surgerydat, theta = log(2))
\end{verbatim}

By default, the control limit is determined on a simulated sample of \(200\) in-control hospitals. The BK-CUSUM is not very computationally expensive. It is therefore usually possible to determine the control limit with higher precision by increasing the \texttt{n\_sim} argument.

\hypertarget{constructing-the-chart}{%
\subsubsection{Constructing the chart}\label{constructing-the-chart}}

We construct the BK-CUSUM on the data of the ninth hospital aiming to detect a doubling in the hazard rate of patients (\texttt{theta} = log(2)).

\begin{verbatim}
BK <- bk_cusum(data = subset(surgerydat, unit == 9), theta = log(2),
               coxphmod = coxph_risk_model)
plot(BK, h = BK_control$h) + ggtitle("BK-CUSUM of hospital 9")
\end{verbatim}

\begin{figure}
\centering
\includegraphics{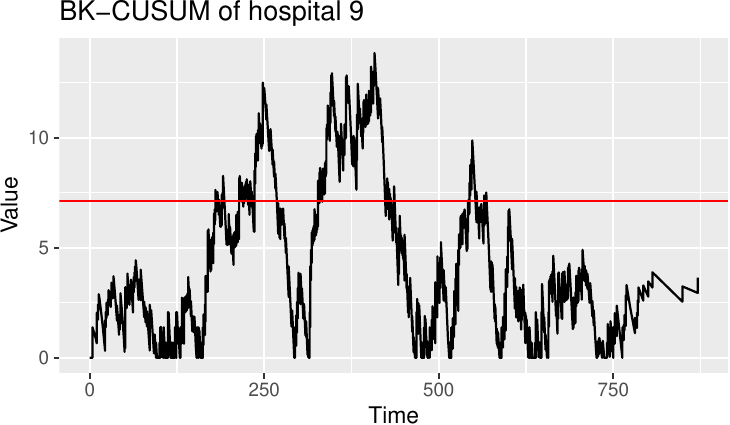}
\caption{\label{fig:bkcusum9}BK-CUSUM for hospital 9 over the study duration. Target performance estimated from the failure rate of patients treated in the first year of the study.}
\end{figure}

The resulting plot is presented in Figure \ref{fig:bkcusum9}. The run length of the chart, visible as the time at which the chart first crosses the red line, can be found using the \texttt{runlength} function.

When the cumulative baseline hazard is not specified through the argument \texttt{cbaseh}, but a Cox Risk-adjustment model \texttt{coxphmod} as obtained from \texttt{coxph} is provided, the cumulative baseline hazard will automatically be determined from this Cox model. When the argument \texttt{ctimes} is left empty, the chart will only be determined at the times of patient failures, as this is sufficient for detection purposes and saves computation time. A control limit \texttt{h} can be specified, so that the chart is only constructed until the value of the chart exceeds the value of the control limit. This is very convenient when monitoring the quality of care at a hospital. Sometimes it is desirable to only construct the chart up until a certain time point, for this the argument \texttt{stoptime} can be used. The argument \texttt{C} can be used to only consider patient outcomes up until \(C\) time units after their surgery. Originally the BK-CUSUM (Biswas and Kalbfleisch 2008) was proposed with \(C = 365\), considering patient outcomes only until \(1\) year post surgery. Finally, a progress bar can be added using the argument \texttt{pb}.

Suppose a decrease in the failure rate is of interest, we can then construct a lower-sided BK-CUSUM by specifying a \texttt{theta} value smaller than \(0\). For example, for detecting a halving of the hazard rate we can take \(\theta = - \log(2)\), such that \(e^\theta = \frac{1}{2}\).

\begin{verbatim}
BK_control_lower <- bk_control_limit(
  time = 365, alpha = 0.05, psi = 1, coxphmod = coxph_risk_model,
  baseline_data = surgerydat, theta = -log(2))
\end{verbatim}

\begin{verbatim}
BK_control_lower <- bk_control_limit(
  time = 365, alpha = 0.05, psi = 1, coxphmod = coxph_risk_model,
  baseline_data = surgerydat, theta = -log(2))
BKlower <- bk_cusum(data = subset(surgerydat, unit == 9), 
              theta = -log(2), coxphmod = coxph_risk_model)
plot(BKlower, h = BK_control_lower$h) + 
  ggtitle("BK-CUSUM of hospital 9 (lower sided)")
\end{verbatim}

\begin{figure}
\centering
\includegraphics{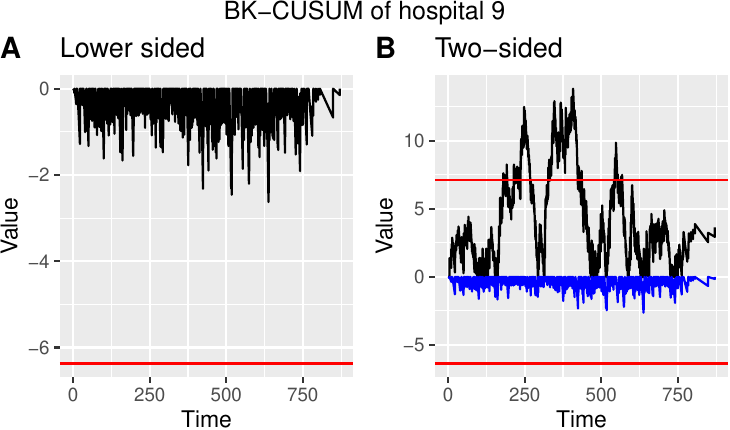}
\caption{\label{fig:bkupperlower}Lower sided (left) and two-sided (right) Bernoulli CUSUM for hospital 9 over the study duration. Target performance estimated from the failure rate of patients treated in the first year of the study.}
\end{figure}

The resulting plot can be found in Figure \ref{fig:bkupperlower} A.

Similarly, when both an increase and decrease of the failure rate are of interest the argument \code{twosided = TRUE} can be used. This produces a two-sided BK-CUSUM (see Figure \ref{fig:bkupperlower} B). For the lower-sided BK-CUSUM the control limit must be determined separately.

\begin{verbatim}
BKtwosided <- bk_cusum(data = subset(surgerydat, unit == 9), 
              theta = log(2), coxphmod = coxph_risk_model, twosided = TRUE)
plot(BKtwosided, h = c(BK_control_lower$h, BK_control$h))
\end{verbatim}

\hypertarget{sec:CGRCUSUMfunction}{%
\subsection{The CGR-CUSUM function}\label{sec:CGRCUSUMfunction}}

The \texttt{cgr\_cusum} function can be used to construct the CGR-CUSUM detailed in Section \protect\hyperlink{sec:CGRCUSUM}{Continuous time Generalized Rapid response CUSUM (CGR-CUSUM)}. This function has almost the same syntax as the \texttt{bk\_cusum} function. The difference is that the procedure estimates a suitable value for \texttt{theta} through maximum likelihood estimation, instead of requiring the users to specify such a value a priori.

The maximum likelihood estimate in the CGR-CUSUM can be unstable at early time points, when not much information is available about subject failure. For this reason, the value of the maximum likelihood estimate is restricted to \(e^{\hat{\theta}} \leq 6\) by default. This comes down to believing that the true hazard ratio at any hospital is always smaller than or equal to \(6\) times the baseline. To change this belief, the user can supply the \texttt{maxtheta} parameter to the \texttt{cgr\_cusum} and \texttt{cgr\_control\_limit} functions.

\hypertarget{determining-control-limits-2}{%
\subsubsection{Determining control limits}\label{determining-control-limits-2}}

Similarly to the \texttt{bk\_control\_limit} function used for the BK-CUSUM, the \texttt{cgr\_control\_limit} function can be used to determine the control limit for the CGR-CUSUM chart as follows:

\begin{verbatim}
CGR_control <- cgr_control_limit(
  time = 365, alpha = 0.05, psi = 1, coxphmod = coxph_risk_model,
  baseline_data = surgerydat)
\end{verbatim}

By default the control limit for the CGR-CUSUM is determined on only \(20\) simulated samples (due to the computational intensity of the procedure), but we recommend to increase the number of samples by using the argument \texttt{n\_sim} to get a more accurate control limit. This will greatly increase the computation time. To speed up the procedure it is possible to parallelize the computations of the CUSUM charts in Step 2 of the simulation procedure (see \protect\hyperlink{sec:controllimitscusum}{Control limits for CUSUM functions}) by specifying the number of cores to use to the argument \texttt{ncores}. Additionally, as the simulation procedure for the CGR-CUSUM can take a lot of time, it is possible to display individual progress bars for each constructed CUSUM chart by specifying \texttt{chartpb\ =\ TRUE}.

\hypertarget{constructing-the-chart-1}{%
\subsubsection{Constructing the chart}\label{constructing-the-chart-1}}

The CGR-CUSUM for the ninth hospital is created with:

\begin{verbatim}
CGR <- cgr_cusum(data = subset(surgerydat, unit == 9), 
               coxphmod = coxph_risk_model)
plot(CGR, h = CGR_control$h) + ggtitle("CGR-CUSUM of hospital 9")
\end{verbatim}

\begin{figure}
\centering
\includegraphics{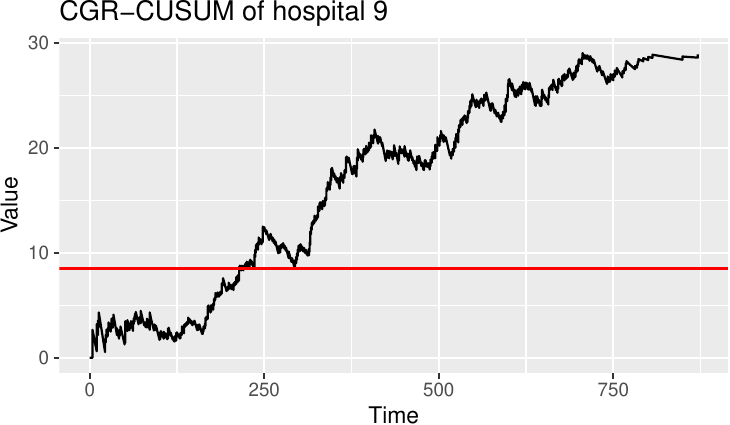}
\caption{\label{fig:cgrcusumfig}CGR-CUSUM for hospital 9 over the study duration. Target performance estimated from the failure rate of patients treated in the first year of the study.}
\end{figure}

The resulting plot can be seen in Figure \ref{fig:cgrcusumfig}. To determine the run length of the procedure (time at which chart crosses the control limit), use the function \texttt{runlength}.

To construct the control chart up until the time of detection, the parameter \(h\) can be specified in the \texttt{cgr\_cusum} function. This allows for continuous inspection, as well as reducing computation time.

Since the CGR-CUSUM is time consuming when the value has to be computed at many time points, we recommend to leave \texttt{ctimes} unspecified, so that the CGR-CUSUM will only be determined at the times necessary for detection purposes.

To reduce computing time, we allow users to parallelize the computations across multiple cores. This can be easily done through the \texttt{ncores} argument. The calculation of the CGR-CUSUM proceeds through 2 steps. First, the contributions to the cumulative intensity of each subject are determined at every time point of interest and are stored in a matrix. Afterwards, the value of the chart is computed by performing matrix operations. When \texttt{ncores\ \textgreater{}\ 1}, both steps are automatically parallelized using functions from the \CRANpkg{pbapply} package (Solymos and Zawadzki 2021). When a value for the control limit has been specified, only the first step can be parallelized. For small data sets and/or short runs \texttt{cmethod\ =\ "CPU"} can be chosen, thereby recalculating the value of the chart at every desired time point but not requiring a lot of initialization. For small hospitals and/or short detection times it could be the preferred method of construction. As the CGR-CUSUM can take long to construct, it is recommended to display a progress bar by specifying \texttt{pb\ =\ TRUE}.

\begin{verbatim}
CGR_multicore <- cgr_cusum(data = subset(surgerydat, unit == 9), 
               coxphmod = coxph_risk_model, ncores = 3, pb = TRUE)
\end{verbatim}

\hypertarget{the-interactive-plot-function}{%
\subsection{The interactive plot function}\label{the-interactive-plot-function}}

The \texttt{interactive\_plot} function can be used to plot multiple CUSUM charts together in one figure, while allowing the user to interact with the plot. This is achieved by using the package \texttt{plotly} (Sievert 2020). We show how to use these features by plotting some of the CUSUM charts from the previous Sections together in one figure. We first combine all CUSUM charts into a list, together with the control limits.

\begin{verbatim}
Bernoulli$h <- bernoulli_control$h
BK$h <- BK_control$h
CGR$h <- CGR_control$h
cusum_list <- list(Bernoulli, BK, CGR)
interactive_plot(cusum_list, unit_names = rep("Hosp 9", 3), scale = TRUE)
\end{verbatim}

\begin{figure}
\centering
\includegraphics{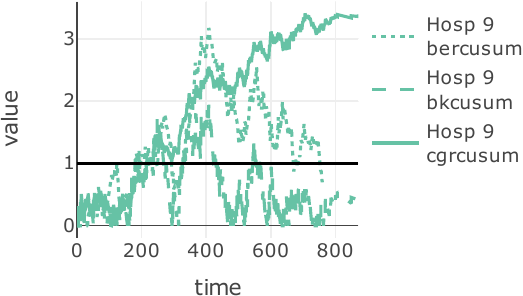}
\caption{\label{fig:interplot-static}Interactive plot of CUSUM charts for hospital 9.}
\end{figure}

As charts have different control limits, it is preferable to scale their values by their respective control limits by specifying \texttt{scale\ =\ TRUE}. After scaling, the control limit will be \(h = 1\) for all CUSUM charts. The resulting plot can be seen in Figure \ref{fig:interplot-static}. By choosing \texttt{highlight\ =\ TRUE}, the user can highlight CUSUM charts by hovering over them. The \CRANpkg{plotly} package allows for many interactive capabilities with the plot.

\hypertarget{sec:application}{%
\section{Application}\label{sec:application}}

In Section \protect\hyperlink{sec:Rpkgsuccess}{The R package success} we employed a simulated data set to show how to use the \CRANpkg{success} package. In this Section, we illustrate the use of the \CRANpkg{success} package on a data set based on a clinical trial for breast cancer conducted by the European Organisation for Research and Treatment of Cancer (EORTC). Covariates for \(2663\) patients over \(15\) treatment centres are available, with patients having surgery over a span of \(61\) time units. In addition, the chronological time of surgery and time since surgery until a combined endpoint are known.

To analyse the data with the \CRANpkg{success} package, we first arrange them in the format presented in Section \protect\hyperlink{sec:Rpkgsuccess}{The R package success}. The outcome of interest is event-free survival. For patients who did not experience an event during the study period, the observations were censored at the last time the patients were known to be event-free. We consider the start of the study as the time when the first patient had surgery. The resulting data was stored in a \texttt{data.frame} called \texttt{breast} which can be loaded using the \texttt{data} function.

To determine the risk-adjustment models, we consider \(36\) time units post surgery followup as outcome. We fit a logistic model to be used for risk adjustment in the funnel plot and Bernoulli CUSUM, and a Cox model for risk adjustment in the BK- and CGR-CUSUM:

\begin{verbatim}
glmmodEORTC <- glm((survtime <= 36) & (censorid == 1) ~ var1 + var2 +
                      var3 + var4 + var5 + var6 + var7,
                   data = breast, family = binomial)
phmodEORTC <- coxph(Surv(survtime, censorid) ~ var1 + var2 +
                      var3 + var4 + var5 + var6 + var7, data = breast)
\end{verbatim}

We then construct the funnel plot and Bernoulli, BK- and CGR-CUSUM for each of the \(15\) centres in the data. To make the continuous time charts visually interesting, we determine their values at every time unit from the start of the study using the argument \texttt{ctimes}.

We then estimate the Poisson arrival rate for each center in the data using the \texttt{arrival\_rate} function.

\begin{verbatim}
arrival_rate <- arrival_rate(breast)
arrival_rate
\end{verbatim}

\begin{verbatim}
#>          1          2          3          5          4          6          7 
#>  0.1568693  0.7127849  0.7922508  0.9657104  0.9676291  1.3067964  1.3870253 
#>          8          9         10         11         12         13         14 
#>  1.4589466  1.6650555  3.0589682  3.6197173  6.8752701  7.5733827 11.0912301 
#>         15 
#> 18.2107083
\end{verbatim}

Based on the estimated arrival rate \(\hat{\psi}\) (per time unit), we group the centres into \(3\) categories:

\begin{itemize}
\tightlist
\item
  Small: Centres 1-5: \(\hat{\psi} \approx 0.9\);
\item
  Medium: Centres 6-11: \(\hat{\psi} \approx 2.1\);
\item
  Large: Centres 12-15: \(\hat{\psi} \approx 11\);
\end{itemize}

For each category, we determine the control limits to use in the CUSUM charts, using the \texttt{*\_control\_limit} functions. For this, we restrict the simulated type I error over \(60\) time units to \(0.05\).

\begin{verbatim}
h <- matrix(0, nrow = 3, ncol = 3, 
            dimnames = list(c(0.9, 2.1, 11), c("Ber", "BK", "CGR")))
psi <- c(0.9, 2.1, 11)
for(i in 1:3){
  h[i,1] <- bernoulli_control_limit(time = 60, alpha = 0.05, followup = 36,
            psi = psi[i], n_sim = 300, theta = log(2), glmmod =  glmmodEORTC,
            baseline_data = breast)$h
  h[i,2] <- bk_control_limit(time = 60, alpha = 0.05, psi = psi[i], n_sim = 300,
            theta = log(2), coxphmod = phmodEORTC, baseline_data = breast)$h
  h[i,3] <- cgr_control_limit(time = 60, alpha = 0.05, psi = psi[i],
            n_sim = 300, coxphmod = phmodEORTC, baseline_data = breast)$h
}
\end{verbatim}

The resulting control limits can be found in Table \ref{tab:htable-static}.

\begin{table}

\caption{\label{tab:htable-static}Control limits for the EORTC data. Estimated using a simulation procedure so that the probability of a type I error within 36 time units is approximately 0.05.}
\centering
\begin{tabular}[t]{l|r|r|r}
\hline
  & Ber & BK & CGR\\
\hline
0.9 & 2.29 & 3.23 & 4.90\\
\hline
2.1 & 2.93 & 3.89 & 5.51\\
\hline
11 & 4.71 & 6.43 & 6.49\\
\hline
\end{tabular}
\end{table}

The columns represent the chart and the rows represent the estimated arrival rate. Using these control limits, we determine the times of detection for the \(15\) centres using the \texttt{runlength} function.

\begin{verbatim}
times_detection <- matrix(0, nrow = 3, ncol = 15, 
                          dimnames = list(c("Ber", "BK", "CGR"), 1:15))
for(i in 1:5){
  times_detection[1,i] <- runlength(EORTC_charts[[i]]$ber, h = h[1,1])
  times_detection[2,i] <- runlength(EORTC_charts[[i]]$bk, h = h[1,2])
  times_detection[3,i] <- runlength(EORTC_charts[[i]]$cgr, h = h[1,3])
}
for(i in 6:11){
  times_detection[1,i] <- runlength(EORTC_charts[[i]]$ber, h = h[2,1])
  times_detection[2,i] <- runlength(EORTC_charts[[i]]$bk, h = h[2,2])
  times_detection[3,i] <- runlength(EORTC_charts[[i]]$cgr, h = h[2,3])
}
for(i in 12:15){
  times_detection[1,i] <- runlength(EORTC_charts[[i]]$ber, h = h[3,1])
  times_detection[2,i] <- runlength(EORTC_charts[[i]]$bk, h = h[3,2])
  times_detection[3,i] <- runlength(EORTC_charts[[i]]$cgr, h = h[3,3])
}
\end{verbatim}

We determine the centres which were detected by any of the charts, and compare their detection times.

\begin{verbatim}
ceiling(times_detection[,colSums(is.infinite(times_detection)) != 3])
\end{verbatim}

\begin{verbatim}
#>       3   5   9 10 11  14
#> Ber  50 Inf  41 45 47 Inf
#> BK  Inf  92 Inf 21 25 127
#> CGR  21 112 Inf 16 23 Inf
\end{verbatim}

The columns represent the centre numbers, while the rows represent the CUSUM charts. We find that the detections by the considered charts do not coincide perfectly with the centres detected by the funnel plot (10, 11) at a \(5\) percent significance level. Comparing the continuous time methods, we see that centre \(5\) is detected faster by the BK-CUSUM, while centres \(10\) and \(11\) are signaled faster by the CGR-CUSUM. Centre \(14\) is only detected by the BK-CUSUM while centre \(9\) is only detected by the Bernoulli CUSUM.

An important consideration when comparing detection times between discrete and continuous time methods is that the discrete time charts inspect the survival probability of patients after \(36\) time units, while the continuous time charts inspect overall survival. This means that the Bernoulli CUSUM might detect a centre with high post operative failure proportions in the \(36\) time units after surgery. However, it does not mean that patients necessarily experience failures faster than expected at this centre as the Bernoulli CUSUM makes no distinction between a patient who has failed \(1\) time unit or \(10\) time units post treatment. This could explain why only the Bernoulli CUSUM detects centre \(9\).

For the continuous time charts it is important to keep in mind that multiple consecutive failures cause the BK-CUSUM to jump up by \(\log(2)\) for every failure, independent of the probability of failure of the patients at that point in time. This can lead to fast detections when many failures are clustered. The CGR-CUSUM can make smaller or larger jumps, depending on the failure probability for each patient at the time of death. This could explain why only the BK-CUSUM detects Centre \(14\). For centres with low volumes of patients, the maximum likelihood estimate in the CGR-CUSUM might not converge quickly to an appropriate value therefore causing a delay in detection times. In contrast, a wrong choice of \(\theta_1\) in the BK-CUSUM may negatively influence detection times (Gomon et al. 2022).

We take a closer look at the disparities between detection times by visualising the funnel plot as well as CUSUM charts for all centres in the EORTC data.

\begin{verbatim}
unnames <- paste(rep("Centre", 15), 1:15)
ber_EORTC <- lapply(EORTC_charts, FUN = function(x) x$ber)
bk_EORTC <- lapply(EORTC_charts, FUN = function(x) x$bk)
cgr_EORTC <- lapply(EORTC_charts, FUN = function(x) x$cgr)
for(i in 1:5){
  ber_EORTC[[i]]$h <- h[1,1] 
  bk_EORTC[[i]]$h <- h[1,2]
  cgr_EORTC[[i]]$h <- h[1,3]
}
for(i in 6:11){
  ber_EORTC[[i]]$h <- h[2,1] 
  bk_EORTC[[i]]$h <- h[2,2]
  cgr_EORTC[[i]]$h <- h[2,3]  
}
for(i in 12:15){
  ber_EORTC[[i]]$h <- h[3,1] 
  bk_EORTC[[i]]$h <- h[3,2]
  cgr_EORTC[[i]]$h <- h[3,3]  
}
cols <- palette.colors(6, "Set2")
col_manual <- rep("lightgrey", 15)
col_manual[c(3,5,9,10,11,14)] <- cols
t1 <- interactive_plot(ber_EORTC, unit_names = unnames, 
                       scale = TRUE, group_by = "type",
                       manual_colors = col_manual)
t2 <- interactive_plot(bk_EORTC, unit_names = unnames, 
                       scale = TRUE, group_by = "type",
                       manual_colors = col_manual)
t3 <- layout(interactive_plot(cgr_EORTC, unit_names = unnames, 
                       scale = TRUE, group_by = "type",
                       manual_colors = col_manual))
t0 <- ggplotly(plot(EORTC_funnel))
\end{verbatim}

\begin{figure}
\centering
\includegraphics{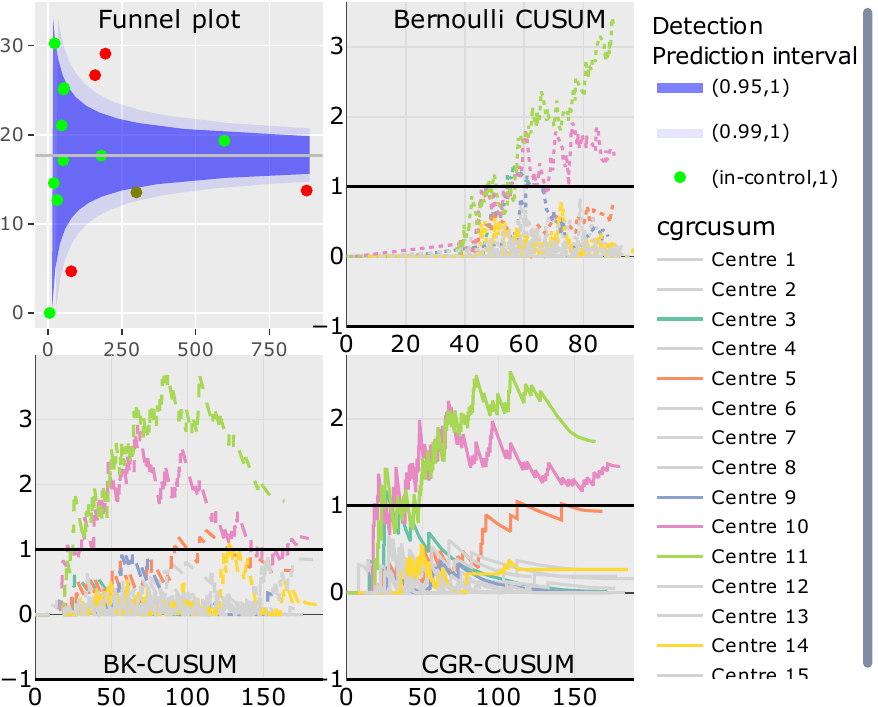}
\caption{\label{fig:funnelandcusums-static}(Top left) Funnel plot of breast data. (Top right) Bernoulli, (Bottom left) BK- and (Bottom right) CGR-CUSUM charts of all 15 centres. Centres not detected by any of the charts are greyed out.}
\end{figure}

The resulting plots are shown in Figure \ref{fig:funnelandcusums-static}.

\begin{figure}
\centering
\includegraphics{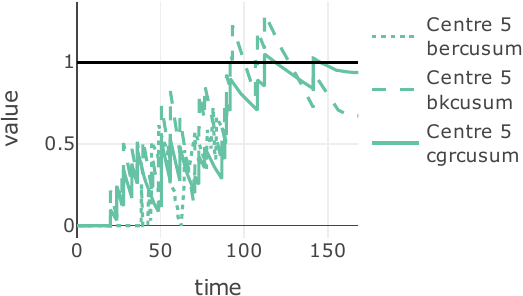}
\caption{\label{fig:EORTC3-static}Bernoulli, BK- and CGR-CUSUM for centre 5 of the breast data.}
\end{figure}

All hospitals detected during the time of the study by the CUSUM charts are highlighted, the remaining centres are shown in gray. We can clearly see the \(36\) time unit delay in the Bernoulli CUSUM charts. The CGR-CUSUM charts have high initial spikes when the first failures are observed. This happens due to the instability of the maximum likelihood estimate \(\hat{\theta}(t)\) when only few failures have been observed. After a sufficient amount of patients have been observed, the CGR-CUSUM makes a clear distinction between centres with respect to their performance. Part of the centres retain a value close to zero while for others the value increases over time. In contrast, the BK-CUSUM charts appear to be less stable, with centres almost hitting the control limit over the period of the study multiple times. This could indicate that the value of \(\theta = \ln(2)\) is not suitable for these centres. Only centres \(10\) (purple) and \(11\) (lightgreen) were detected by all charts. From the value of the continuous time charts we can presume that centre \(10\) had a cluster of failures at the beginning of the study, followed by a period of (slightly) above average failures. The Bernoulli CUSUM does not provide such insights, as failures come in \(36\) time units after surgery. It seems that centre \(11\) had a high rate of failure from the start until the end of the study. Centre \(9\) (dark blue) was only detected by the Bernoulli CUSUM. This disparity between discrete and continuous time charts mostly happens when the failure proportions at \(36\) time units post surgery are relatively large, but many patients fail at reasonable times (e.g.~around \(30\) time units post surgery). From a visual inspection of the charts, this seems to be the case for Centre \(9\). Finally, Centre \(5\) was only detected by the continuous time CUSUM charts. We display all \(3\) CUSUM charts for Centre \(5\) in Figure \ref{fig:EORTC3-static}. The Bernoulli CUSUM only incorporates the information provided by survival \(36\) time units after surgery. Because of this, the Bernoulli CUSUM can only be calculated up to \(36\) time units after the last patient had surgery (in this case, until the \(89\)th time unit). The continuous time charts can incorporate failures of patients at any point in time, therefore producing signals at later times. While the BK-CUSUM always rises by \(\ln(2)\) whenever a failure is observed, the CGR-CUSUM can make jumps of different sizes, depending on the risk of failure of the observed patient. This causes a disparity in the times of detection and also in interpretation of the values of the charts. As no patients had surgery later than \(60\) time units after the start of the study, the arrival rate after \(60\) time units is \(\psi = 0\) for all centres, meaning detections at that point should be taken with a grain of salt.

\hypertarget{discussion}{%
\section{Discussion}\label{discussion}}

The \CRANpkg{success} package implements three CUSUM methods for the inspection of the failure rate in survival data in continuous time and the funnel plot. Using the \texttt{parameter\_assist} function, quality control charts can also be constructed by users unfamiliar with control chart and survival theory.

We would like to highlight the different type of outcome and purpose of the control charts in this package. The funnel plot should not be used for the continuous inspection of survival data, as it can only be used to test for a difference in failure proportion at a fixed point in time. The Bernoulli CUSUM is closest to the funnel plot, as it uses the same outcome to determine chart values. The Bernoulli CUSUM is suitable for continuous inspection, but the followup time has a great impact on the resulting conclusions as well as the choice of the expected increase in hazard ratio \texttt{theta}. In contrast, the BK-CUSUM can incorporate patient failures at any point in time, but also requires the specification of \texttt{theta} a priori. The CGR-CUSUM does not have this downside, and as the increase in failure rate is never known in advance in practical applications, it can lead to quicker detection times. Finally, the CUSUM charts test different hypotheses, nuancing their interpretation even further: the Bernoulli and CGR-CUSUM charts can be used to test for a change in the failure rate starting from some patient, while the BK-CUSUM can be used to test for a sudden change in the failure rate of all patients.

The funnel plot, Bernoulli CUSUM and BK-CUSUM do not require a lot of computational power, whereas the computation of the CGR-CUSUM is more sophisticated and can require more time. For this reason, we provide the user the option to parallelize the computation of this chart. A key part of CUSUM charts are their control limits, which are mostly determined using simulation studies due to the lack of analytical results. This can be done using the \texttt{*\_control\_limit} functions in the \CRANpkg{success} package. The time required to compute control limits depends on the value of the arrival rate \texttt{psi} and the proportion of failures in the data. A higher value of \texttt{psi} means more patients have to be accounted for, and a higher failure proportion means chart values need to be calculated more often. For the breast cancer data, determining control limits took approximately \(10\) minutes in total (using a consumer grade laptop), on a simulated sample of \(300\) in-control centres.

It is important to determine appropriate control limits for the inspection of survival processes using CUSUM charts. The chosen value of \texttt{psi} greatly influences the value of the control limit. Heuristically, this can be compared with the prediction intervals in the funnel plot. As the number of outcomes in a centre increases, the prediction intervals become narrower. For the CUSUM charts, this is expressed in the control limit.

\hypertarget{acknowledgements}{%
\section{Acknowledgements}\label{acknowledgements}}

The authors thank the European Organization for Research and Treatment of Cancer for permission to apply our methods on data based on an EORTC study. The contents of this publication and methods used are solely the responsibility of the authors and do not necessarily represent the official views of the EORTC.

\hypertarget{references}{%
\section*{References}\label{references}}
\addcontentsline{toc}{section}{References}

\hypertarget{refs}{}
\begin{CSLReferences}{1}{0}
\leavevmode\vadjust pre{\hypertarget{ref-qichartsR}{}}%
Anhoej, Jacob. 2021. \emph{{qicharts}: Quality Improvement Charts}. \url{https://CRAN.R-project.org/package=qicharts}.

\leavevmode\vadjust pre{\hypertarget{ref-austin_all}{}}%
Austin, P. C. 2012. {``Generating Survival Times to Simulate {Cox} Proportional Hazards Models with Time-Varying Covariates.''} \emph{Statistics in Medicine} 31 (29): 3946--58. \url{https://doi.org/10.1002/sim.5452}.

\leavevmode\vadjust pre{\hypertarget{ref-biswas_kalbfleisch}{}}%
Biswas, P., and J. D. Kalbfleisch. 2008. {``A Risk‐adjusted CUSUM in Continuous Time Based on the {C}ox Model.''} \emph{Statistics in Medicine} 27: 3452--52. \url{https://doi.org/10.1002/sim.3216}.

\leavevmode\vadjust pre{\hypertarget{ref-cook_all}{}}%
Cook, David, M Coory, and R Webster. 2011. {``Exponentially Weighted Moving Average Charts to Compare Observed and Expected Values for Monitoring Risk-Adjusted Hospital Indicators.''} \emph{BMJ Quality and Safety} 20 (May): 469--74. \url{http://dx.doi.org/10.1136/bmjqs.2008.031831}.

\leavevmode\vadjust pre{\hypertarget{ref-cox}{}}%
Cox, D. R. 1972. {``Regression Models and Life-Tables.''} \emph{Journal of the Royal Statistical Society. Series B (Methodological)} 34 (2): 187--220. \url{https://doi.org/10.1111/j.2517-6161.1972.tb00899.x}.

\leavevmode\vadjust pre{\hypertarget{ref-qcrR}{}}%
Flores, Miguel. 2021. \emph{{qcr}: Quality Control Review}. \url{https://CRAN.R-project.org/package=qcr}.

\leavevmode\vadjust pre{\hypertarget{ref-Gomon_all_2022}{}}%
Gomon, Daniel, Hein Putter, Rob G. H. H. Nelissen, and Stéphanie van der Pas. 2022. {``CGR-Cusum: A Continuous Time Generalized Rapid Response Cumulative Sum Chart.''} \emph{Biostatistics}. \url{https://doi.org/10.1093/biostatistics/kxac041}.

\leavevmode\vadjust pre{\hypertarget{ref-ggQCR}{}}%
Grey, Kenith. 2018. \emph{ggQC: Quality Control Charts for 'Ggplot'}. \url{https://CRAN.R-project.org/package=ggQC}.

\leavevmode\vadjust pre{\hypertarget{ref-grigg}{}}%
Grigg, O. A. 2018. {``The {STRAND} Chart: A Survival Time Control Chart.''} \emph{Statistics in Medicine} 38 (9): 1651--61. \url{https://doi.org/10.1002/sim.8065}.

\leavevmode\vadjust pre{\hypertarget{ref-cusumR}{}}%
Hubig, Lena. 2019. \emph{{cusum}: Cumulative Sum (CUSUM) Charts for Monitoring of Hospital Performance}. \url{https://CRAN.R-project.org/package=cusum}.

\leavevmode\vadjust pre{\hypertarget{ref-keefe_all}{}}%
Keefe, Matthew J., Justin B. Loda, Ahmad E. Elhabashy, and William H. Woodall. 2017. {``Improved Implementation of the Risk-Adjusted Bernoulli CUSUM Chart to Monitor Surgical Outcome Quality.''} \emph{International Journal for Quality in Health Care} 29 (3): 343--48. \url{https://doi.org/10.1093/intqhc/mzx036}.

\leavevmode\vadjust pre{\hypertarget{ref-spcR}{}}%
Knoth, Sven. 2021. \emph{{spc}: Statistical Process Control -- Calculation of ARL and Other Control Chart Performance Measures}. \url{https://CRAN.R-project.org/package=spc}.

\leavevmode\vadjust pre{\hypertarget{ref-funnelR}{}}%
Kumar, Matthew. 2018. \emph{funnelR: Funnel Plots for Proportion Data}. \url{https://CRAN.R-project.org/package=funnelR}.

\leavevmode\vadjust pre{\hypertarget{ref-qccR}{}}%
Scrucca, Luca. 2004. {``Qcc: An {R} Package for Quality Control Charting and Statistical Process Control.''} \emph{R News} 4/1: 11--17. \url{https://CRAN.R-project.org/package=qcc}.

\leavevmode\vadjust pre{\hypertarget{ref-plotlyR}{}}%
Sievert, Carson. 2020. \emph{Interactive Web-Based Data Visualization with {R}, Plotly, and Shiny}. Chapman; Hall/CRC. \url{https://plotly-r.com}.

\leavevmode\vadjust pre{\hypertarget{ref-pbapplyR}{}}%
Solymos, Peter, and Zygmunt Zawadzki. 2021. \emph{{pbapply}: Adding Progress Bar to '*Apply' Functions}. \url{https://CRAN.R-project.org/package=pbapply}.

\leavevmode\vadjust pre{\hypertarget{ref-spiegelhalter}{}}%
Spiegelhalter, D. J. 2005. {``Funnel Plots for Comparing Institutional Performance.''} \emph{Statistics in Medicine} 24 (8): 1185--1202. \url{https://doi.org/10.1002/sim.1970}.

\leavevmode\vadjust pre{\hypertarget{ref-steiner2000}{}}%
Steiner, S. H., R. J. Cook, V. T. Farewell, and T. Treasure. 2000. {``Monitoring Surgical Performance Using Risk-Adjusted Cumulative Sum Charts.''} \emph{Biostatistics} 1 (4): 441--52. \url{https://doi.org/10.1093/biostatistics/1.4.441}.

\leavevmode\vadjust pre{\hypertarget{ref-steiner_jones}{}}%
Steiner, S. H., and M. Jones. 2009. {``Risk-Adjusted Survival Time Monitoring with an Updating Exponentially Weighted Moving Average {(EWMA)} Control Chart.''} \emph{Statistics in Medicine} 29 (4): 444--54. \url{https://doi.org/10.1002/sim.3788}.

\leavevmode\vadjust pre{\hypertarget{ref-survivalR}{}}%
Terry M. Therneau, and Patricia M. Grambsch. 2000. \emph{Modeling Survival Data: Extending the {C}ox Model}. New York: Springer.

\leavevmode\vadjust pre{\hypertarget{ref-ggplot2R}{}}%
Wickham, Hadley. 2016. \emph{{ggplot2}: Elegant Graphics for Data Analysis}. Springer-Verlag New York. \url{https://ggplot2.tidyverse.org}.

\leavevmode\vadjust pre{\hypertarget{ref-vladR}{}}%
Wittenberg, Philipp, and Sven Knoth. 2020. \emph{{vlad}: Variable Life Adjusted Display and Other Risk-Adjusted Quality Control Charts}. \url{https://CRAN.R-project.org/package=vlad}.

\end{CSLReferences}

\bibliography{RJwrapper.bib}

\address{%
Daniel Gomon\\
Leiden University\\%
Mathematical Institute\\ Niels Bohrweg 1\\ 2333CA Leiden, the Netherlands\\
\url{https://github.com/d-gomon}\\%
\textit{ORCiD: \href{https://orcid.org/0000-0001-9011-3743}{0000-0001-9011-3743}}\\%
\href{mailto:d.gomon@math.leidenuniv.nl}{\nolinkurl{d.gomon@math.leidenuniv.nl}}%
}

\address{%
Marta Fiocco\\
Leiden University\\%
Mathematical Institute\\ Niels Bohrweg 1\\ 2333CA Leiden, the Netherlands\\
\textit{ORCiD: \href{https://orcid.org/0000-0001-5588-0277}{0000-0001-5588-0277}}\\%
}

\address{%
Hein Putter\\
Leiden University\\%
Mathematical Institute\\ Niels Bohrweg 1\\ 2333CA Leiden, the Netherlands\\
\textit{ORCiD: \href{https://orcid.org/0000-0001-5395-1422}{0000-0001-5395-1422}}\\%
}

\address{%
Mirko Signorelli\\
Leiden University\\%
Mathematical Institute\\ Niels Bohrweg 1\\ 2333CA Leiden, the Netherlands\\
\textit{ORCiD: \href{https://orcid.org/0000-0002-8102-3356}{0000-0002-8102-3356}}\\%
}

\end{article}

\end{document}